\documentclass[aip, numerical, amsmath, amssymb, jcp]{revtex4-1}

\usepackage{graphicx}
\usepackage{dcolumn}
\usepackage{bm}

\usepackage[utf8]{inputenc}
\usepackage[T1]{fontenc}
\usepackage{mathptmx}
\usepackage{color}

\begin{document}

\title{Time-dependent Density Matrix Renormalization Group Quantum Dynamics for Realistic Chemical Systems}

\author{Xiaoyu Xie}
\affiliation{School of Chemistry and Chemical Engineering, Nanjing University, Nanjing 210023, China}
\author{Yuyang Liu}
\affiliation{School of Chemistry and Chemical Engineering, Nanjing University, Nanjing 210023, China}
\author{Yao Yao}
\affiliation{Department of Physics and State Key Laboratory of Luminescent Materials and Devices, South China University of Technology, Guangzhou 510640, China}
\author{Ulrich Schollw\"{o}ck}
\email{schollwoeck@lmu.de}
\affiliation{Department of Physics, Arnold Sommerfeld Center for Theoretical Physics (ASC), Fakult\"{a}t f\"{u}r Physik, Ludwig-Maximilians-Universit\"{a}t M\"{u}nchen, M\"{u}nchen D-80333, Germany}
\affiliation{Munich Center for Quantum Science and Technology (MCQST), Schellingstr. 4, M\"{u}nchen D-80799, Germany}
\author{Chungen Liu}
\email{cgliu@nju.edu.cn}
\affiliation{School of Chemistry and Chemical Engineering, Nanjing University, Nanjing 210023, China}
\author{Haibo Ma}
\email{haibo@nju.edu.cn}
\affiliation{School of Chemistry and Chemical Engineering, Nanjing University, Nanjing 210023, China}

\date{\today}

\begin{abstract}
Electronic and/or vibronic coherence has been found by recent ultrafast spectroscopy experiments in many chemical, biological, and material systems. This indicates that there are strong and complicated interactions between electronic states and vibration modes in realistic chemical systems. Therefore, simulations of quantum dynamics with a large number of electronic and vibrational degrees of freedom are highly desirable. Due to the efficient compression and localized representation of quantum states in the matrix-product state (MPS) formulation, time-evolution methods based on the MPS framework, which we summarily refer to as tDMRG (time-dependent density-matrix renormalization group) methods, are considered to be promising candidates to study the quantum dynamics of realistic chemical systems. In this work, we benchmark the performances of four different tDMRG methods, including global Taylor, global Krylov, and local one-site and two-site time-dependent variational principles (1TDVP and 2TDVP), with a comparison to multiconfiguration time-dependent Hartree and experimental results. Two typical chemical systems of internal conversion and singlet fission are investigated: one containing strong and high-order local and nonlocal electron-vibration couplings and the other exhibiting a continuous phonon bath. The comparison shows that the tDMRG methods (particularly, the 2TDVP method) can describe the full quantum dynamics in large chemical systems accurately and efficiently. Several key parameters in the tDMRG calculation including the truncation error threshold, time interval, and ordering of local sites were also investigated to strike the balance between efficiency and accuracy of results.
\end{abstract}

\maketitle

\section{Introduction} \label{sec: 1}
Solving the time-(in)dependent Schr\"{o}dinger equation of a given nonrelativistic quantum system is the most straightforward and commonly pursued idea to study its static properties and out-of-equilibrium behavior. Unfortunately, it is almost impossible to obtain exact solutions for large systems as the dimension of the configuration space grows exponentially with increasing system sizes. To tackle this so-called curse of dimensionality, many theoretical methods using different approximations have been proposed. Among them, tensor product methods \cite{kolda2009tensor, dolgov2014tensor} recently attracted a lot of research interest. They treat quantum states and operators, expressed in products of local bases, as high-order tensors with an exponentially increasing number of coefficients and decompose these high-order tensors by different algorithms into suitable products of many low-rank and localized low-order tensors to compress the wavefunction and reduce the computational costs. One well-known decomposition algorithm is Tucker decomposition \cite{tucker1966some} used in multiconfiguration time-dependent Hartree (MCTDH) \cite{meyer1990multi, beck2000multiconfiguration}, which decomposes a high-order tensor with high rank into a set of matrices and one small Tucker core tensor with the same order but low rank; it can be considered as a highorder single value decomposition (HOSVD) \cite{de2000best, de2000multilinear}. For higher orders, the Tucker core still suffers from the curse of dimensionality. This can be overcome by introducing multilayer MCTDH (ML-MCTDH) \cite{wang2003multilayer} using a hierarchical Tucker (HT) decomposition. The tensor train \cite{oseledets2011tensor} (TT, in the mathematical literature) decomposition or the equivalent matrix product state (MPS, in the physical literature) representation \cite{verstraete2008matrix, schollwock2011density} used in the density matrix renormalization group (DMRG) \cite{white1992density, schollwock2005density} provides an alternative decomposition algorithm, which decomposes a high-order tensor with high rank into a product of many local low-order tensors with a one-dimension (1D) topology. This decomposition method has been generalized to tensor network states (TNS) such as projected entangled-pair states (PEPS) \cite{verstraete2004renormalization, verstraete2008matrix} or tensor tree networks (TTN) \cite{shi2006classical, nakatani2013efficient, gunst2018t3ns, schroder2019tensor} for non-1D systems.

Among various tensor product methods, the DMRG has been widely recognized as the most accurate numerical tool for calculating 1D strongly correlated systems, because of its two advantages of an efficient compression and a localized structure of its underlying MPS/TT formulation of the wavefunction. Following the success of the DMRG in describing equilibrium quantities, various time-dependent variants (which we globally refer to as a tDMRG) have been developed over the last 15 years, extending the MPS/TT/DMRG to explore the real-time quantum dynamics of strongly correlated systems by solving the time-dependent Schrödinger equation (TDSE). In 2004, White and Feiguin \cite{white2004real}, Daley \textit{et al.} \cite{daley2004time} and Verstraete \textit{et al.} \cite{verstraete2004matrix} proposed algorithms based on the time-evolving block decimation (TEBD) algorithm of Vidal \cite{vidal2003efficient, vidal2004efficient}. They all used a Suzuki-Trotter decomposition \cite{suzuki1976generalized} of the Hamiltonian into its individual two-body terms. When these terms are local, their time-evolution propagation operator can be applied efficiently to the wavefunction in the MPS format. Unfortunately, Trotterization is not easily applicable to Hamiltonians with long-range interactions, which appear in quasi-2D systems or quantum chemistry applications as well as system-bath problems, but only after the use of numerous time-consuming site-swapping operations (“swap gates”). In order to deal with such general Hamiltonians, one can implement global time integration solvers (e.g. Runge-Kutta \cite{ronca2017time, ren2018time, frahm2019ultrafast} and Krylov \cite{paeckel2019time} approaches) for the TDSE based on the compressed MPS wavefunction directly for the tDMRG without explicitly constructing the time evolution propagator. To exploit the second advantage of a localized structure in the MPS/TT/DMRG, local tDMRG methods, including local Krylov \cite{garcia2006time, paeckel2019time} and time-dependent variational principle (TDVP) \cite{haegeman2011time, lubich2015time, haegeman2016unifying} approaches, were also developed, which solve a sequence of localized effective differential equations for timeevolution by introducing appropriate projectors of MPS/TT and inserting the projectors into the original TDSE. Moreover, the MPS/TT structure in the tDMRG has also been successfully utilized in describing the mean-field operators in MCTDH, \cite{kurashige2018matrix, benedikt2019multiset}, the reduced/auxiliary density operators or density vectors in hierarchical equations of motion (HEOM) \cite{shi2018efficient, borrelli2019density} simulations, and the gridbased wavepacket in split-operator Fourier transform (SOFT) \cite{greene2017tensor} very recently.

In the last few years, tDMRG methods have been applied successfully for realistic chemical systems with electron-vibration (electron-phonon) interactions. In 2017, Borrelli and co-workers \cite{borrelli2017simulation} studied the large-scale exciton quantum dynamics using TDVP methods for the Fenna-Matthews-Olsen complex, which has more than 500 vibrational degrees of freedom and 7 electronic sites. In this work, the finite temperature effect was also examined by their developed methodology based on thermofield dynamics (TFD) theory \cite{borrelli2016quantum} which combines an accurate description of quantum dynamics at finite temperature with the flexibility of a basis set representation, in a similar spirit to the thermofield-based chain-mapping approach proposed by de Vega and Ba{\~n}uls \cite{de2015thermofield}. TDVP methods were also applied by Borrelli and co-workers to study electron-transfer problems in realistic models comprising more than 200 nuclear vibrations coupled to the electronic states \cite{borrelli2018theoretical} and the coherences in a model excitonic system \cite{gelin2019origin}. In 2018, some of us implemented a unitary transformation approach for realistic vibronic Hamiltonians and simulated the charge transfer dynamics and 2D electronic spectrum of the oligothiophene/fullerene interface in organic solar cells via TEBD methods. \cite{yao2018full} Shuai and co-workers simulated the absorption and fluorescence spectra of perylene bisimide (PBI) and distyrylbenzene molecular aggregates at zero and finite temperature by tDMRG methods using the fourth-order Runge-Kutta method combined with the TFD approach. \cite{ren2018time} In 2019, the TDVP methods were used by Reiher and Baiardi \cite{baiardi2019large} for the real- and imaginary-time evolution of PBI, pyrazine, and ethylene systems containing a large number of degrees of freedom. Recently, the TDVP approach was optimized by Ren and co-workers \cite{li2019numerical} with the help of graphical processing units (GPUs) and multicore central processing units (CPUs); calculations of excitation energy transfer in FMO systems were found to be accelerated by a factor of up to 57 by using GPUs. These studies suggest the great potential of tDMRG methods for exploring the quantum dynamics of large chemical systems.

In order to provide practical guidelines for the future applications of various tDMRG methods in simulating the realistic chemical systems, in this work, we benchmark four different tDMRG methods (global Taylor, global Krylov, and one-site and two-site TDVPs) for two different chemical systems and compare our results with MCTDH, ML-MCTDH, and experimental results. In Section~\ref{sec: 2}, we give a brief introduction to the framework of MPS and matrix product operators (MPO) (Section~\ref{sec: 2.1.1}), global and local tDMRG methods (Section~\ref{sec: 2.1.2}), and a general model for the study of chemical systems with electronic-vibration/phonon interactions (Section~\ref{sec: 2.2}). The computational details and calculation results of S$_1$/S$_2$ internal conversion in pyrazine and singlet fission (SF) in molecular dimer are given in Section~\ref{sec: 3.1} and \ref{sec: 3.2} respectively. Section~\ref{sec: 4} presents the summary of this work.

\section{Methodology} \label{sec: 2}
As the details of DMRG and tDMRG have been discussed elsewhere \cite{schollwock2011density, ma2018time, paeckel2019time}, we only briefly introduce the basic ideas of tDMRG methods in section~\ref{sec: 2.1} and show how they are implemented for the simulation of the quantum dynamics of realistic chemical system by introducing models for electron-vibration interaction in section~\ref{sec: 2.2}.
\subsection{tDMRG methods} \label{sec: 2.1}
\subsubsection{MPS and MPO} \label{sec: 2.1.1}
A general quantum state with $n$ local sites can be expressed as a linear combination of orthonormal configurations (in terms of the $d_i$-dimensional local basis $\{\vert\sigma_i\rangle\}$ ):
\begin{equation} \label{eq: linear comb}
\vert\psi\rangle=\sum_{\{\sigma_i\}}c_{\sigma_1\sigma_2\cdots\sigma_n}\vert\sigma_1\sigma_2\cdots\sigma_n\rangle.
\end{equation}
The set of coefficient $c_{\sigma_1\sigma_2\cdots\sigma_n}$ can be regarded as a high-dimension tensor, which can be decomposed to the tensor train (TT) structure. \cite{oseledets2011tensor} In physics language, the state can be reformulated as an MPS:
\begin{eqnarray}
\vert\psi\rangle&=&\sum_{\{\sigma_i\},\{\alpha_i\}}A^{\sigma_1}_{1,\alpha_1}A^{\sigma_2}_{\alpha_1,\alpha_2}\cdots A^{\sigma_n}_{\alpha_{n-1},1}\vert\sigma_1\sigma_2\cdots\sigma_n\rangle, \label{eq: MPS tensor}\\
&=&\sum_{\{\sigma_i\}}\mathbf{A}^{\sigma_1}\mathbf{A}^{\sigma_2}\cdots \mathbf{A}^{\sigma_n}\vert\sigma_1\sigma_2\cdots\sigma_n\rangle. \label{eq: MPS matrix}
\end{eqnarray}
Each $\alpha_i$ is summed from 1 to $m_i$. The rank-3 tensor $A^{\sigma_i}_{\alpha_{i-1}, \alpha_i}$ ($\alpha_0=\alpha_n=1$) has two bond legs ($\alpha_{i-1}$, $\alpha_i$) and one physical leg ($\sigma_i$), and the bond dimension $m=\max{(m_i)}$ is related to the amount of entanglement in $\vert\psi\rangle$ and is decreased by various truncation procedures such as singular value decomposition (SVD). Practically, the truncation can be performed either by fixing the maximal number $M$ of retained singular values or dynamical block state selection (DBSS) \cite{legeza2003controlling} via discarding singular values below a fixed threshold $\varepsilon$. Compared to the fixed $M$ approach, the (t)DMRG accuracy by using DBSS method is more stable for systems with different sizes and/or coupling strengths. \cite{legeza2003controlling} Using controls of constant truncation rather than constant bond dimension has become standard practice for at least a decade because fixed bond dimensions lead to strongly varying accuracy as time evolves, which makes extrapolation to the formally exact limit of infinite bond dimension very difficult. Constant truncation ensures a smoother evolution of the errors and a drastically better extrapolation behavior to the formally exact limit of zero truncation (which is then the same as infinite bond dimension). In this work, we use the DBSS method for MPS compression in all our tDMRG simulations
unless otherwise stated. $\mathbf{A}^{\sigma_i}$ is a rank-2 subtensor (matrix) of $A^{\sigma_i}_{\alpha_{i-1},\alpha_i}$ with two indexes $\alpha_{i-1}$ (row index) and $\alpha_i$ (column index) for each $\vert\sigma_i\rangle$ of local site $i$.

In particular, a tensor $L^{\sigma_i}_{\alpha_{i-1},\alpha_i}$ ($R^{\sigma_i}_{\alpha_{i-1},\alpha_i}$) is called a left (right) orthonormal tensor if it satisfies the following equations,
\begin{eqnarray}
\sum_{\sigma_i,\alpha_{i-1}}L^{\sigma_i}_{\alpha_{i-1},\alpha_i}(L^{\sigma_i}_{\alpha_{i-1},\alpha_i^\prime})^*&=&\delta_{\alpha_i\alpha_i^\prime}, \label{eq: l orthonormal} \\
\sum_{\sigma_i,\alpha_i}(R^{\sigma_i}_{\alpha_{i-1},\alpha_i})^*R^{\sigma_i}_{\alpha_{i-1}^\prime,\alpha_i}&=&\delta_{\alpha_{i-1}\alpha_{i-1}^\prime}. \label{eq: r orthonormal}
\end{eqnarray}
Due to the gauge transformation symmetry ($A^{\sigma_i}\rightarrow A^{\sigma_i}X^{-1}$, $A^{\sigma_{i+1}}\rightarrow XA^{\sigma_{i+1}}$) of MPS, each MPS can be rebuilt as a left (right) canonical MPS consisting only of left (right) orthonormal tensor components or a so-called mixed-canonical MPS,
\begin{eqnarray}
\vert\psi\rangle&=&\sum_{\{\sigma_k\}}\mathbf{L}^{\sigma_1}\mathbf{L}^{\sigma_2}\cdots\mathbf{M}^{\sigma_i}\cdots\mathbf{R}^{\sigma_{n-1}}\mathbf{R}^{\sigma_n}\vert\sigma_1\sigma_2\cdots\sigma_n\rangle, \\
&=&\sum_{\sigma_i, \alpha_{i-1},\alpha_i}M^{\sigma_i}_{\alpha_{i-1},\alpha_i}\vert{\cal L}^{[1:i-1]}_{\alpha_{i-1}}\rangle\vert\sigma_i\rangle\vert{\cal R}^{[i+1:n]}_{\alpha_i}\rangle \label{eq: mixed MPS}
\end{eqnarray}
$\vert{\cal L}^{[1:i-1]}_{\alpha_{i-1}}\rangle$ ($\vert{\cal R}^{[i+1:n]}_{\alpha_i}\rangle$) are block configurations with the left (right) orthonormal basis, the site $i$ with arbitrary tensor component $M^{\sigma_i}_{\alpha_{i-1}, \alpha_i}$ is called active site or orthogonality center.

Correspondingly, every operator expressed in the local basis set $\{\vert\sigma_1\cdots\sigma_n\rangle\}$ can be rewritten as an MPO,
\begin{equation} \label{eq: MPO}
\hat{O}=\sum_{\{\sigma_i\},\{\sigma_i^\prime\},\{w_i\}}W^{\sigma_1,\sigma_1^\prime}_{1,w_1}\cdots W^{\sigma_n,\sigma_n^\prime}_{w_{n-1},1}\vert\sigma_1\cdots\sigma_n\rangle\langle\sigma_1^\prime\cdots\sigma_n^\prime\vert,
\end{equation}
where the local tensor component $W^{\sigma_i,\sigma_i^\prime}_{w_{i-1},w_i}$ ($w_1=w_n=1$) is a rank-4 tensor with two bond legs ($w_{i-1}$, $w_i$) and two physical legs ($\sigma_i$, $\sigma_i^\prime$) acting on ket and bra states, respectively. By analogy with the definition of bond dimension of MPS, the bond dimension of an MPO can be defined as $m_{\mathrm{MPO}}=\max{(w_i)}$. Particularly, the bond dimension of Hamiltonian provide information of the complexity for describing both static properties and dynamic behaviour of system.

For the framework of MPS, two operations are key, the overlap between two states (MPS) and the application of an operator (MPO) to a state (MPS).

The calculation of the overlap between two MPS is straightforward due to the orthonormality of the basis,
\begin{eqnarray}
\langle\psi^\prime\vert\psi\rangle&=&\sum_{\{\sigma^\prime_i\},\{\alpha^\prime_i\}}\left(A^{\prime\sigma^\prime_1}_{1,\alpha^\prime_1}A^{\prime\sigma^\prime_1}_{\alpha^\prime_1,\alpha^\prime_2}\cdots A^{\prime\sigma^\prime_n}_{\alpha^\prime_{n-1},1}\right)^\dagger\langle\sigma^\prime_1\sigma^\prime_2\cdots\sigma^\prime_n\vert \nonumber \\
&&\times\sum_{\{\sigma_i\},\{\alpha_i\}}A^{\sigma_1}_{1,\alpha_1}A^{\sigma_2}_{\alpha_1,\alpha_2}\cdots A^{\sigma_n}_{\alpha_{n-1},1}\vert\sigma_1\sigma_2\cdots\sigma_n\rangle \label{eq: overlap 1} \\
&=&\sum_{\{\sigma_i\}}(\mathbf{A}^{\prime\sigma_n})^\dagger\cdots(\mathbf{A}^{\prime\sigma_1})^\dagger\mathbf{A}^{\sigma_1}\cdots\mathbf{A}^{\sigma_n}. \label{eq: overlap 2}
\end{eqnarray}
For an efficient evaluation, the final expression has to be evaluated in a suitable order; see e.g. \cite{schollwock2011density}.

The application of an MPO to an MPS can be carried out directly or variationally.
\begin{enumerate}
\item The direct procedure consists of straightforward tensor contractions between the local components of the MPS and the MPO,
\begin{eqnarray}
\hat{O}\vert\psi\rangle&=&\sum_{\{\sigma_i\},\{\sigma_i^\prime\},\{w_i\}}W^{\sigma_1,\sigma_1^\prime}_{1,w_1}\cdots W^{\sigma_n,\sigma_n^\prime}_{w_{n-1},1}\vert\sigma_1\cdots\sigma_n\rangle\langle\sigma_1^\prime\cdots\sigma_n^\prime\vert \nonumber \\
&&\times \sum_{\{\sigma_i^{\prime\prime}\},\{\alpha_i\}}A^{\sigma_1^{\prime\prime}}_{1,\alpha_1}\cdots A^{\sigma_n^{\prime\prime}}_{\alpha_{n-1},1}\vert\sigma_1^{\prime\prime}\cdots\sigma_n^{\prime\prime}\rangle \label{eq: direct app 1} \\[0.5 cm]
&=&\sum_{\{\sigma_i\},\{w_i, \alpha_i\}}A^{\prime\sigma_1}_{1,w_1\alpha_1}\cdots A^{\prime\sigma_n}_{w_{n-1}\alpha_{n-1},1}\vert\sigma_1\cdots\sigma_n\rangle, \label{eq: direct app 2}
\end{eqnarray}
with the new tensor components $A^{\prime\sigma_i}_{w_{i-1}\alpha_{i-1},w_i\alpha_i}$ given by
\begin{equation}
A^{\prime\sigma_i}_{w_{i-1}\alpha_{i-1},w_i\alpha_i}=\sum_{\sigma^\prime_i}W^{\sigma_i\sigma^\prime_i}_{w_{i-1},w_i}A^{\sigma^\prime_i}_{\alpha_{i-1},\alpha_i}
\end{equation}
\item The variational procedure approximates the result by minimizing the distance between the sought-for resulting MPS $\vert\phi\rangle$ and the MPO applied to the initial MPS,
\begin{equation} \label{eq: variational app}
\min_{\vert\phi\rangle}\Vert\vert\phi\rangle-\hat{O}\vert\psi\rangle\Vert^2.
\end{equation}
By optimizing each local component $A^{\prime\sigma_i}_{\alpha_{i-1},\alpha_i}$ of $\vert\phi\rangle$ and sweeping sites until convergence, the minimization in~(\ref{eq: variational app}) is achieved, resulting in $\vert\phi\rangle$.
\end{enumerate}

\subsubsection{Time-evolution in the MPS framework: tDMRG methods} \label{sec: 2.1.2}
To solve the time-dependent Schr\"{o}dinger equation (TDSE) and perform the time-evolution of a quantum state $\vert\psi(t)\rangle$, we can define the propagation operator $\hat{U}(t,t+\delta t)$,
\begin{equation} \label{eq: propagator}
\hat{U}(t,t+\delta t)=e^{-i\hat{H}(t)\delta t/\hbar},
\end{equation}
such that the state is updated in time as
\begin{equation} \label{eq: time evolution}
\vert\psi(t+\delta t)\rangle = \hat{U}(t, t+\delta t)\vert\psi(t)\rangle.
\end{equation}
There are many ways to solve the TDSE by constructing the propagator $\hat{U}$ (such as by a Taylor expansion, by diagonalizing the Hamiltonian, or by Chebyshev methods \cite{tal1984accurate}) or by updating the state without explicit construction of $\hat{U}$ (e.g. Runge-Kutta methods and Krylov subspace methods \cite{nevanlinna2012convergence, saad2003iterative}). \cite{moler2003nineteen} tDMRG methods combine these time integrators with the advantages of the MPS framework (such as the efficient truncation of quantum states and separate treatment of local sites).

If we ignore the special localized structure of the MPS/MPO representation and apply time evolution to the compressed MPS wavefunction directly, several global time-integration solvers can be implemented easily. For instance, the MPS can be updated via the 4th-order Runge-Kutta method (for time-independent $\hat{H}$),
\begin{eqnarray}
\vert K_1\rangle&=&-\frac{i\delta t}{\hbar}\hat{H}\vert\psi(t)\rangle, \label{eq: runge kutta 1} \\
\vert K_2\rangle&=&-\frac{i\delta t}{\hbar}\hat{H}(\vert\psi(t)\rangle+\frac{1}{2}\vert K_1\rangle), \label{eq: runge kutta 2} \\
\vert K_3\rangle&=&-\frac{i\delta t}{\hbar}\hat{H}(\vert\psi(t)\rangle+\frac{1}{2}\vert K_2\rangle), \label{eq: runge kutta 3} \\
\vert K_4\rangle&=&-\frac{i\delta t}{\hbar}\hat{H}(\vert\psi(t)\rangle+\vert K_3\rangle), \label{eq: runge kutta 4} \\
\vert\psi(t+\delta t)\rangle&=&\vert\psi(t)\rangle+\frac{1}{6}(\vert K_1\rangle+2\vert K_2\rangle+2\vert K_3\rangle+\vert K_4\rangle). \label{eq: runge kutta}
\end{eqnarray}

The Krylov subspace methods were also applied for the timeevolution of MPS, namely, the global Krylov method (for a detailed discussion, see Ref. \cite{paeckel2019time}). The order-$r$ Krylov subspace generated by a Hamiltonian $\hat{H}$ and an initial state $\vert\psi\rangle$ is the linear subspace spanned by images of $\vert\psi\rangle$ under the first $r$ power of $\hat{H}$,
\begin{equation} \label{eq: Krylov subspace}
{\cal K}_r(\hat{H}, \vert\psi\rangle)=\mathrm{span}\{\vert\psi\rangle,\hat{H}\vert\psi\rangle,\cdots,\hat{H}^r\vert\psi\rangle\}.
\end{equation}
After orthonormalization of the subspace in Eq.~\ref{eq: Krylov subspace}, we could compute the exact $\hat{U}^\prime$ in this subspace and the projector $\hat{P}_r$ onto orthonormalized ${\cal K}_r$, and the updated MPS is given by
\begin{equation} \label{eq: global Krylov}
\vert\psi(t+\delta t)\rangle\simeq\hat{P}_r^\dagger\hat{U}^\prime\hat{P}_r\vert\psi(t)\rangle.
\end{equation}

The global methods have the advantage of nearly exactly representing the operation of high-order Hamiltonian as $\hat{H}^n\vert\psi\rangle$, however they may become inefficient for large complicated systems because of the much larger entanglement of the intermediate states (e.g. $\vert K_i\rangle$ in Eq.~\ref{eq: runge kutta 1}-\ref{eq: runge kutta 4} and $\hat{H}^i\vert\psi\rangle$ in Eq.~\ref{eq: Krylov subspace}), which requires a much larger bond dimension. To overcome this bottleneck, the second advantage of the MPS framework (the existence of localised tensor components of MPS and MPO) can be put to use by considering local time evolution steps.

The Trotter-based methods \cite{vidal2003efficient, vidal2004efficient, white2004real, daley2004time, verstraete2004matrix} are among the tDMRG methods working locally by splitting the total Hamiltonian $\hat{H}$ into localised terms $\hat{H}_1$, $\hat{H}_2$,..., $\hat{H}_n$. Then the first-order TEBD propagation operator is expressed as
\begin{equation} \label{eq: TEBD1}
\hat{U}(t,t+\delta t)=e^{-i\sum_j\hat{H}_j(t)\delta t/\hbar}\simeq\Pi_je^{-i\hat{H}_j(t)\delta t/\hbar}.
\end{equation}
Because of the local characteristic of each $\hat{H}_j$, the MPO form of $e^{-i\hat{H}_j(t)\delta t/\hbar}$ can be constructed easily. Trotter-based methods are very efficient for short range interaction systems due to the small number of terms, while it is not efficient for systems containing many long range interaction terms.

Another important local method is the TDVP method \cite{haegeman2011time, haegeman2016unifying, lubich2015time}. The basic idea of the TDVP method is very simple and based on
\begin{equation} \label{eq: Dirac-Frenkel}
\min_{\vert\psi(t)\rangle}\Vert\hat{H}\vert\psi(t)\rangle-i\hbar\frac{\partial}{\partial t}\vert\psi(t)\rangle\Vert^2,
\end{equation}
which is the variational version of TDSE (the Dirac-Frenkel principle); ${\cal F}(t)=\Vert\hat{H}\vert\psi(t)\rangle-i\hbar\frac{\partial}{\partial t}\vert\psi(t)\rangle\Vert^2$ is the Dirac-Frenkel functional.

With the MPS framework, the problem in Eq.~\ref{eq: Dirac-Frenkel} gets reduced to the minimization of Dirac-Frenkel functional ${\cal F}(t)$ with respect to the tensor components of the MPS. This can be achieved by projecting $\hat{H}\vert\psi(t)\rangle$ onto the tangent space of the given $\vert\psi(t)\rangle$ in the tensor manifold. The projector on the tangent space $\hat{P}_{T, \vert\psi(t)\rangle}$ is defined as
\begin{equation} \label{eq: TDVP projector}
\hat{P}_{T,\vert\psi(t)\rangle}=\sum_{i=1}^n\hat{P}^L_{i-1}\otimes\hat{I}_i\otimes\hat{P}^R_{i+1}-\sum_{i=1}^{n-1}\hat{P}^L_{i}\otimes\hat{P}^R_{i+1},
\end{equation}
where $\hat{P}^L_i$ ($\hat{P}^R_i$) is the left (right) block projector,
\begin{eqnarray}
\hat{P}^L_i&=&\sum_{\alpha_i}\vert{\cal L}_{\alpha_i}^{[1:i]}\rangle\langle{\cal L}_{\alpha_i}^{[1:i]}\vert, \label{eq: left projector}\\
\hat{P}^R_i&=&\sum_{\alpha_i}\vert{\cal R}_{\alpha_{i-1}}^{[i:n]}\rangle\langle{\cal R}_{\alpha_{i-1}}^{[i:n]}\vert. \label{eq: right projector}
\end{eqnarray}
The original TDSE is then rewritten in a local version by inserting the projector (Eq.~\ref{eq: TDVP projector}) on both sides of the original TDSE,
\begin{equation} \label{eq: TDVP TDSE}
i\hbar\frac{\partial}{\partial t}\vert\psi(t)\rangle=\hat{P}_{T,\vert\psi(t)\rangle}\hat{H}\vert\psi(t)\rangle.
\end{equation}
Eq~\ref{eq: TDVP TDSE} can be solved approximately by solving $n$ forward-evolving equations,
\begin{equation} \label{eq: forward TDVP}
i\hbar\frac{\partial}{\partial t}\vert\psi(t)\rangle=\sum_{i=1}^n\hat{P}^L_{i-1}\otimes\hat{1}_i\otimes\hat{P}^R_{i+1}\hat{H}\vert\psi(t)\rangle,
\end{equation}
and $n-1$ backward-evolving equations,
\begin{equation} \label{eq: backward TDVP}
i\hbar\frac{\partial}{\partial t}\vert\psi(t)\rangle=-\sum_{i=1}^{n-1}\hat{P}^L_{i}\otimes\hat{P}^R_{i+1}\hat{H}\vert\psi(t)\rangle.
\end{equation}

In practice, Eq.\ref{eq: forward TDVP} and \ref{eq: backward TDVP} are solved in the sequence defined by the order of the local sites. By considering the wavefunction structure of an MPS in Eq.\ref{eq: mixed MPS}, it is easy to find out that the $i$-th term of Eq.\ref{eq: forward TDVP} can be integrated with a time-differential equation as follows,
\begin{equation} \label{eq: one-site TDSE}
i\hbar\dot{\mathbf{M}}^{\sigma_i}=\sum_{\sigma^\prime}\hat{H}_\mathrm{eff,1}^{\sigma_i,\sigma_i^\prime}\mathbf{M}^{\sigma_i^\prime},
\end{equation}
where $\hat{H}_\mathrm{eff,1}^{\sigma_i,\sigma_i^\prime}=\langle\sigma_i\vert\hat{H}_\mathrm{eff, 1}^{i}\vert\sigma_i^\prime\rangle$, in which $\hat{H}_\mathrm{eff,1}^{i}$ is the one-site effective Hamiltonian and defined as
\begin{equation} \label{eq: one-site Hamiltonian 1}
\hat{H}_\mathrm{eff,1}^{i}=\langle{\cal R}_{\alpha_i}^{[i+1:n]}\vert\otimes\langle{\cal L}_{\alpha_{i-1}}^{[1:i-1]}\vert\hat{H}\vert{\cal L}_{\alpha_{i-1}^\prime}^{[1:i-1]}\rangle\otimes\vert{\cal R}_{\alpha_i^\prime}^{[i+1:n]}\rangle,
\end{equation}
and it has two physical legs ($\sigma_i$, $\sigma_i^\prime$) and four bond legs ($\alpha_{i-1}$, $\alpha_i$, $\alpha_{i-1}^\prime$ and $\alpha_i^\prime$). For each term in Eq.\ref{eq: backward TDVP}, a similar time dependent equation can be derived by introducing zero-site effective Hamiltonian as
\begin{equation} \label{eq: zero-site Hamiltonian}
\hat{H}_\mathrm{eff,0}^{i}=\langle{\cal R}_{\alpha_{i,r}}^{[i+1:n]}\vert\otimes\langle{\cal L}_{\alpha_{i,l}}^{[1:i]}\vert\hat{H}\vert{\cal L}_{\alpha_{i,l}^\prime}^{[1:i]}\rangle\otimes\vert{\cal R}_{\alpha_{i,r}^\prime}^{[i+1:n]}\rangle,
\end{equation}
which has only four bond legs ($\alpha_{i,l}$, $\alpha_{i,r}$, $\alpha_{1,l}^\prime$ and $\alpha_{i,r}^\prime$).

With these definitions, one can achieve a simple scheme for one-site TDVP (1TDVP) algorithm: We start from $i=1$ with a right-canonical form of the initial MPS, evolve the tensor component $M^{\sigma_i}_{\alpha_{i-1},\alpha_i}$ based on Eq.~\ref{eq: one-site TDSE}, then left-orthonormalize the updated $\mathbf{M}^{\sigma_i}=\mathbf{L}^{\sigma_i}\mathbf{Q}^i$, evolve the zero-site $Q^i_{\alpha_i,\alpha_i}$ backward  using zero-site effective Hamiltonian $\hat{H}_\mathrm{eff,0}^{i}$ before absorbing it into the next tensor component $\mathbf{M}^{\sigma_{i+1}}=\mathbf{Q}^i\mathbf{R}^{\sigma_{i+1}}$. We then repeat these steps for $i=i+1$ to complete a left-to-right sweeping procedure. An interesting property of 1TDVP is that the projection of the Hamiltonian onto the MPS manifold occurs before the time evolution and no truncation has to happen after the evolution. As such, both the norm and energy of the state are conserved under real-time evolution. However, because 1TDVP is developed for the purpose of constraining the time evolution to a specific manifold of MPS of a given initial bond dimension, the bond dimension of MPS does not increase in 1TDVP during the time evolution, which is contradictory to the fact that the quantum entanglement is increasing for the real-time propagated MPS, which requires an enlarged bond dimension. In this case, we performed the global method (Krylov method in our case) in the initial several steps before the 1TDVP calculation for constructing an MPS with a sufficiently large bond dimension to reduce the error of later 1TDVP simulations.

Another solution is to use a two-site approach, namely, 2TDVP method, instead of 1TDVP. For the 2TDVP approach, a two-site effective Hamiltonian is used for forward-evolving parts,
\begin{equation} \label{eq: two-site Hamiltonian}
\hat{H}_\mathrm{eff,2}^{i,i+1}=\langle{\cal R}_{\alpha_{i+1}}^{[i+2:n]}\vert\otimes\langle{\cal L}_{\alpha_{i-1}}^{[1:i-1]}\vert\hat{H}\vert{\cal L}_{\alpha_{i-1}^\prime}^{[1:i-1]}\rangle\otimes\vert{\cal R}_{\alpha_{i+1}^\prime}^{[i+2:n]}\rangle.
\end{equation}
The two-site effective Hamiltonian consists of four physical legs ($\sigma_i$, $\sigma_{i+1}$, $\sigma_i^\prime$ and $\sigma_{i+1}^\prime$) and four bond legs ($\alpha_{i-1}$, $\alpha_{i+1}$, $\alpha_{i-1}^\prime$ and $\alpha_{i+1}^\prime$). Moreover, there is information of bond leg of Hamiltonian ($w_i$) in the two-site effective Hamiltonian. Therefore, the bond dimension of MPS in 2TDVP can be adaptively increased, but truncation is necessary after the two-site local tensor component by using SVD to keep the efficiency of the 2TDVP calculation, i.e., 2TDVP has a smaller projection error but a larger truncation error compared to the 1TDVP approach, which has exactly a zero truncation error but a larger projection error. It should be noted that the 2TDVP can evolve Hamiltonians with only nearestneighbor interactions exactly without incurring a projection error. The detailed procedure of 2TDVP is similar to that of the 1TDVP method: Again, we start from $i=1$ with a right-canonical form of the initial MPS, evolve the two-site tensor component $M^{\sigma_i,\sigma_{i+1}}_{\alpha_{i-1},\alpha_{i+1}}=\sum_{\alpha_i}M^{\sigma_i}_{\alpha_{i-1},\alpha_i}R^{\sigma_{i+1}}_{\alpha_{i},\alpha_{i+1}}$ based on a two-site effective Hamiltonian, then left-orthonormalize the updated $\mathbf{M}^{\sigma_i,\sigma_{i+1}}=\mathbf{L}^{\sigma_i}\mathbf{M}^{\sigma_{i+1}}$, evolve the one-site $M^{\sigma_{i+1}}_{\alpha_i,\alpha_{i+1}}$ backward before absorbing it into the next two-site tensor $M^{\sigma_{i+1},\sigma_{i+2}}_{\alpha_{i},\alpha_{i+2}}=\sum_{\alpha_{i+1}}M^{\sigma_{i+1}}_{\alpha_{i},\alpha_{i+1}}R^{\sigma_{i+2}}_{\alpha_{i+1},\alpha_{i+2}}$, update $i=i+1$, and then repeat the steps before. In addition, in order to avoid being trapped in local minima with the initial small bond dimensions in 2TDVP, one can also first perform global methods for a few steps before starting 2TDVP.

\subsection{Electron-vibration interaction model} \label{sec: 2.2}
Electron-vibration (el-vib) interaction models are used widely for the study of the nonadiabatic dynamics of organic molecular systems with ultrafast features and electron-nuclear interaction, e.g. the S$_1$/S$_2$ spectrum issue of pyrazine and the singlet fission (SF) in organic crystalline system, which will be discussed in section~\ref{sec: 3}.

With a second-order truncation, the Hamiltonian for an el-vib coupled system can be expressed as
\begin{equation} \label{eq: Hamiltonian}
\hat{H}=\hat{H}_{el}+\hat{H}_{vib}+\hat{H}_{el-vib},
\end{equation}
with
\begin{equation} \label{eq: Hel_1}
\hat{H}_{el}=\sum_{i}\varepsilon^0_{i}\vert\psi_i\rangle\langle\psi_i\vert+\sum_{i\neq j}V^0_{ij}\vert\psi_i\rangle\langle\psi_j\vert,
\end{equation}
\begin{equation} \label{eq: Hvib_1}
\hat{H}_{vib}=\sum_{I}\frac{1}{2}\hbar\omega_{I}(-\frac{\partial^2}{\partial Q_I^2}+Q_I^2),
\end{equation}
\begin{equation} \label{eq: Helvib_1}
\hat{H}_{el-vib}=\sum_{i,j,I}g^{I}_{ij}Q_I\vert\psi_i\rangle\langle\psi_j\vert + \sum_{i,j,I,J}g^{IJ}_{ij}Q_IQ_J\vert \psi_i\rangle\langle\psi_j\vert .
\end{equation}
Here, $\varepsilon^0_{i}$ and $V^0_{ij}$ represent the energy of the electronic state $\vert \psi_i\rangle$ and the electronic coupling between $\vert \psi_i\rangle$ and $\vert \psi_j\rangle$ under the equilibrium geometry, respectively. $\omega_I$ is the frequency of the vibration mode $I$ while $Q_I$ is the dimensionless displacement. In the el-vib coupling, $g_{ii}^I$ are linear local el-vib couplings and $g_{ij}^I$ ($i \neq j$) are non-local ones; $g^{IJ}_{ij}$ are the 2nd-order terms.

Considering the discrete basis and matrix representation of the Hamiltonian using the MPS/MPO language, the second quantization is convenient for the tDMRG applications, and the three terms can be rewritten as
\begin{equation} \label{eq: Hel_2}
\hat{H}_{el}=\sum_{i}\varepsilon^0_{i}\hat{a}^\dagger_i\hat{a}_i+\sum_{i\neq j}V^0_{ij}\hat{a}^\dagger_i\hat{a}_j,
\end{equation}
\begin{equation} \label{eq: Hvib_2}
\hat{H}_{vib}=\sum_{I}\hbar\omega_{I}(\hat{b}^\dagger_{I}\hat{b}_{I}+\frac{1}{2}),
\end{equation}
\begin{equation} \label{eq: Helvib_2}
\hat{H}_{el-vib}=\sum_{i,j,I}g^{I}_{ij}\hat{q}_I\hat{a}^\dagger_i\hat{a}_j + \sum_{i,j,I}g^{IJ}_{ij}\hat{q}_I\hat{q}_J\hat{a}^\dagger_i\hat{a}_j.
\end{equation}
Here, $\hat{a}_i^\dagger$ ($\hat{a}_i$) and $\hat{b}_I^\dagger$ ($\hat{b}_I$) are the creators (annihilators) of electronic states $\vert\psi_i\rangle$ and vibration mode $I$, respectively. $\hat{q}_I=\frac{1}{\sqrt{2}}(\hat{b}_I^\dagger+\hat{b}_I)$ is the dimensionless displacement operator of vibration mode $I$.

Obviously, the occupation number representation of Bosonic states is the most convenient basis set for vibration sites. As the configuration number of this discrete basis is infinite, an effective basis truncation is necessary for the MPS and MPO construction. In our case, the basis set for a vibration site $I$ is set to be $\{\vert0\rangle_I, \vert1\rangle_I,\cdots,\vert n_{max}\rangle_I\}$, where the maximal occupation number $n_{max}$ is carefully chosen by testing the convergence of dynamics under changes of $n_{max}$.

\section{Results and discussion} \label{sec: 3}
Two chemical systems are investigated via tDMRG methods. The first one is the S$_1$/S$_2$ interconversion dynamics of the pyrazine molecule system where two electronic states are coupled to 24 discrete vibration modes by local, nonlocal, and 2nd-order el-vib couplings. The second one is the singlet fission in a molecular dimer with three electronic states affected by a continuous phonon bath. All tDMRG calculations are performed using the SyTen package, originally created by Claudius Hubig. \cite{hubig:_syten_toolk, hubig17:_symmet_protec_tensor_networ}

\subsection{The S$_1$/S$_2$ dynamics of pyrazine system} \label{sec: 3.1}
As a well-defined benchmark model system, the S$_1$/S$_2$ pyrazine system coupled with 24 molecular vibration modes has been studied by many quantum dynamics methods \cite{seidner1992binitio, woywod1994characterization, worth1996effect, raab1999molecular, burghardt2008multimode, saller2014basis, kloss2017implementation, greene2017tensor, baiardi2019large} for several decades. The potential energy surfaces of singlet states S$_1$ and S$_2$ states of pyrazine have been computed via \emph{ab initio} quantum chemistry methods such as configuration interaction (CI) \cite{raab1999molecular} and complete active space self-consistent field (CASSCF) methods \cite{woywod1994characterization}, which allows us to build the el-vib interaction Hamiltonian to investigate the dynamics of this system,
\begin{eqnarray*} \label{eq: H of pyr}
\hat{H}&=&\left(
\begin{array}{cc}
-\Delta & 0 \\
 0 & \Delta
\end{array}
\right)+\sum_{I = 1}^{24}\hbar\omega_I(\hat{b}_I^\dagger\hat{b}_I+\frac{1}{2}) \\
&&+\sum_{I=1}^{24}\hat{q}_I\left(
\begin{array}{cc}
g_1^{I} & g_{12}^{I} \\
g_{12}^{I} & g_2^{I}
\end{array}
\right) +\sum_{I,J=1}^{24}\hat{q}_I\hat{q}_J\left(
\begin{array}{cc}
g_1^{IJ} & g_{12}^{IJ} \\
g_{12}^{IJ} & g_2^{IJ}
\end{array}
\right).
\end{eqnarray*}
$\Delta=(E_{S_2}-E_{S_1})/2$ represents the electronic terms and the parameters in vibration and el-vib interaction terms can be found in references \cite{worth1996effect, raab1999molecular}.

\begin{figure}
\includegraphics[scale=0.20]{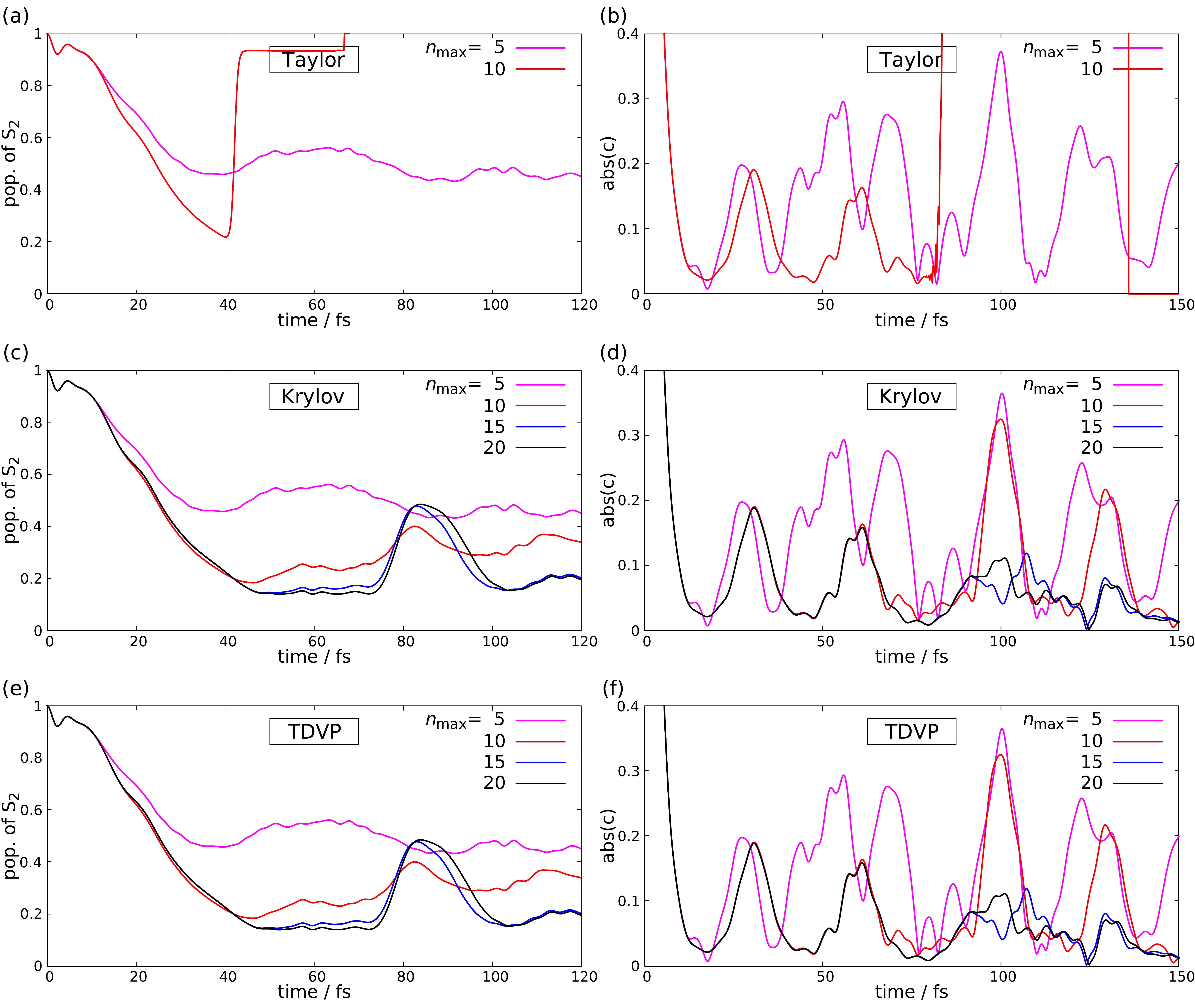}
\caption{The dynamics simulation tests of 4-modes pyrazine S$_1$/S$_2$ systems via (a, b) a 2nd-order Taylor expansion, (c, d) the Krylov method and (e, f) the 2-site TDVP method with different maximal occupation number of vibration sites. (a, c, d) show results of population evolution of S$_2$ and (b, d, f) give results of the absolute value of the autocorrelation function.} \label{fig: pyrazine1}
\end{figure}

First, we test the performance of three tDMRG methods for the simplest 4-modes ($v6a$, $v1$, $v9a$ and $v10$ in reference \cite{worth1996effect}) el-vib interaction models with different maximal occupation numbers $n_{max}$ of vibration sites. The initial state is set as $\vert\psi(0)\rangle=\vert S_2\rangle\otimes\vert0\rangle_{v6a}\otimes\vert0\rangle_{v1}\otimes\vert0\rangle_{v9a}\otimes\vert0\rangle_{v10}$ and total simulation time is fixed as $120$ fs with $5.0$ a.u. for one step. The truncation threshold for SVD during construction of MPS and time-evolution is fixed as $\varepsilon=1 \times 10^{-15}$. The results of electronic population and autocorrelation function $C(t)=\langle\psi(0)\vert\psi(t)\rangle$ are shown in Fig.~\ref{fig: pyrazine1}.

\begin{figure}
\includegraphics[scale=0.20]{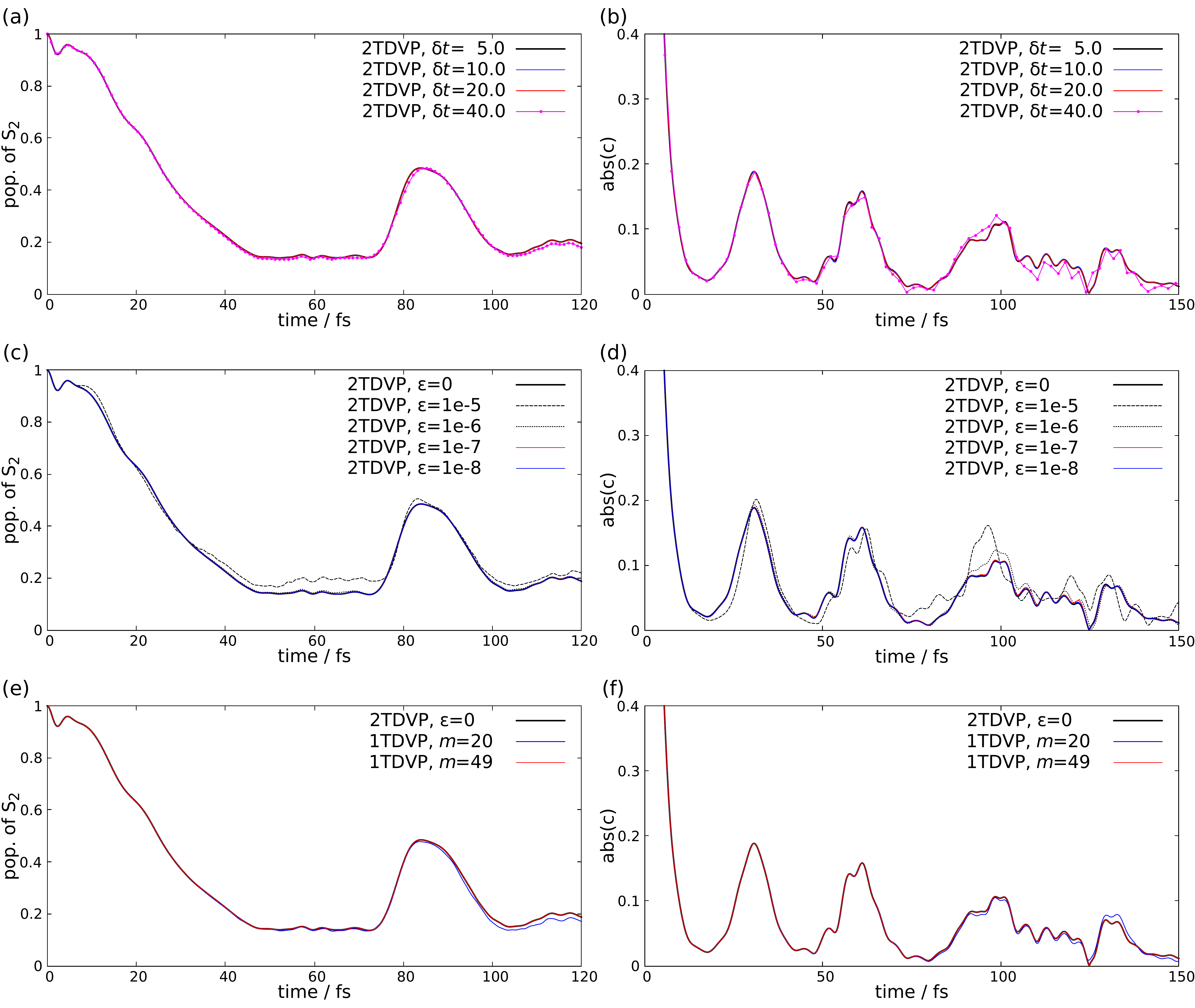}
\caption{The parameters tests of TDVP methods with the dynamics simulation of 4-mode pyrazine S$_1$/S$_2$ systems. (a, b). The tests result of the population evolution of S$_2$ and the absolute value of autocorrelation function with different time intervals ($\delta t=$ 5.0, 10.0, 20.0 and 40.0 a.u. respectively). (c, d). The test results for different truncations $\varepsilon$ in 2TDVP. (e, f). The results for difference initial bond dimension via 1TDVP method.} \label{fig: pyrazine2}
\end{figure}

The results of second-order Taylor expansion in Fig.~\ref{fig: pyrazine1} (a, b) illustrate a crashed simulation with the maximal occupation numbers of 10 for all 4 modes, which suggests that the 2nd-order Taylor expansion method is invalid for large systems due to the large truncation error and unconserved energy and norm. The results of the global Krylov method and the 2TDVP method are consistent with each other, but the 2TDVP method is more stable and less time-consuming than the global Krylov methods. 1TDVP doesn't work properly for this case (Fig~\ref{fig: pyrazine2}) and will be discussed in the next paragraphs. Therefore, the 2TDVP method is applied in the following calculations. On the other hand, the maximal occupation number test in Fig.~\ref{fig: pyrazine1} shows that systems with smaller $n_{max}$ give incorrect dynamics results at earlier times, which demonstrates the increasing relevance of high occupation number states in the basis of Bosonic modes for a faithful representation of the quantum state; relatively large local basis sets for the vibration sites are needed for tDMRG calculation with a given simulation time. We investigated the suitable maximal occupation number for each mode individually to minimize the size of total basis. We fixed the maximal occupation numbers at 24, 18, 10 and 18 for mode $v6a$, $v1$, $v9a$ and $v10$.

With the chosen tDMRG method (the 2TDVP method) and maximal occupation number of each Bosonic site, we also test the different time interval and truncation approximation of the 2TDVP method for the balance of efficiency and accuracy. The results are given in Fig.~\ref{fig: pyrazine2}. The time interval tests in Fig~\ref{fig: pyrazine2} (a, b) display very little variation in the time-evolution behavior for results achieved with $\delta t \leq 20.0$ a.u.; the trajectory for $\delta t = 40.0$ a.u. time intervals show an evident difference after about 50 fs. This suggests that the simulation time interval of pyrazine systems should be no larger than 20.0 a.u. ($\sim 0.48$ fs).

As mentioned in Section~\ref{sec: 2}, the MPS can be truncated effectively to reduce computational cost. The DBSS \cite{legeza2003controlling} approach which fixed threshold $\varepsilon$ ensures a more even quality of simulation results and used here. Fig.~\ref{fig: pyrazine2} (c, d) illustrates the tests for the DBSS approach using 2TDVP method. We noticed from Fig.~\ref{fig: pyrazine2} (c, d) that the convergence of autocorrelation function requires a higher truncation threshold ($\varepsilon=1 \times 10^{-8}$) than the state population ($\varepsilon=1 \times 10^{-7}$), verifying state populations tend to have a quicker convergence than other wavefunction observable like correlation functions. We also tested the performance of 1TDVP method by simulating dynamics as a function of initial bond dimension in Fig.~\ref{fig: pyrazine2}(e, f). Compared to 2TDVP, the accuracy of 1TDVP is highly dependent on the initial bond dimension and the convergence is reached with an initial bond dimension $m=49$. This result indicated that 1TDVP method could be easily implemented for the calculation of relatively small systems, while for large system, a preliminary preparation for a sufficiently large MPS bond dimension is necessary for 1TDVP method.

The ordering of sites is another crucial issue for the performance of DMRG/tDMRG methods, a good ordering may lead to smaller bond dimension of MPO and MPS. We compare the time consumptions for the 4-mode system with seven different orderings by random selection and physical intuition of placing the highly correlated sites as close as possible. The results indicate that a proper ordering $\{v9a, v6a, \vert\mathrm{S}_1\rangle, v10a, \vert\mathrm{S}_2\rangle, v1\}$ based on the magnitude order of the el-vib couplings can be 10 times faster than the worst ordering $\{\vert\mathrm{S}_1\rangle, v9a, v1, v6a, v10a, \vert\mathrm{S}_2\rangle\}$ which puts the two electronic sites with small basis and large entanglement to other phonon modes the two ends of the chain and a default ordering in code as $\{\vert\mathrm{S}_1\rangle, \vert\mathrm{S}_2\rangle, v6a, v10a, v1, v9a\}$ whose electronic states are followed by vibration sites with the ascending order of magnitude of frequency value. The results verify that the ordering of local sites will affect the efficiency of tDMRG calculation significantly and an optimal ordering should arrange sites with large interactions to be close to each other and locate the entangled sites with small basis at the center of the chain. \cite{xiang1996density, legeza2003optimizing} In fact, this may be explained by analyzing the MPS bond dimension of different orderings. In Fig.~\ref{fig: pyrazine4}(c), we plotted dimension of each physical legs ($d_i$) and also the time-dependent dimension of bond legs ($m_i$) for the default and optimized orderings. One may notice that $m_i$ grows more rapidly in the optimized ordering, due to the larger bond dimension of MPO ($m_\mathrm{MPO}=6$ comparing with $m_\mathrm{MPO}=4$ for default ordering). However, the optimal ordering has an advantage of providing a relatively smaller local tensors for each site. It is found that the dimension maximums of both $d_i$ and $m_i$ occur at the same indices 3 and 4 for the default ordering, which will contribute to a large two-site effective tensor with size $m_{i-1}d_i d_{i+1}m_{i+1}$. On the contrary, in the optimal ordering, the maximum of $m_i$ and the minimum of $d_i$ occur at the same index 3, beneficial to reduce the size of largest two-site effective tensor.

\begin{figure}
\includegraphics[scale=0.28]{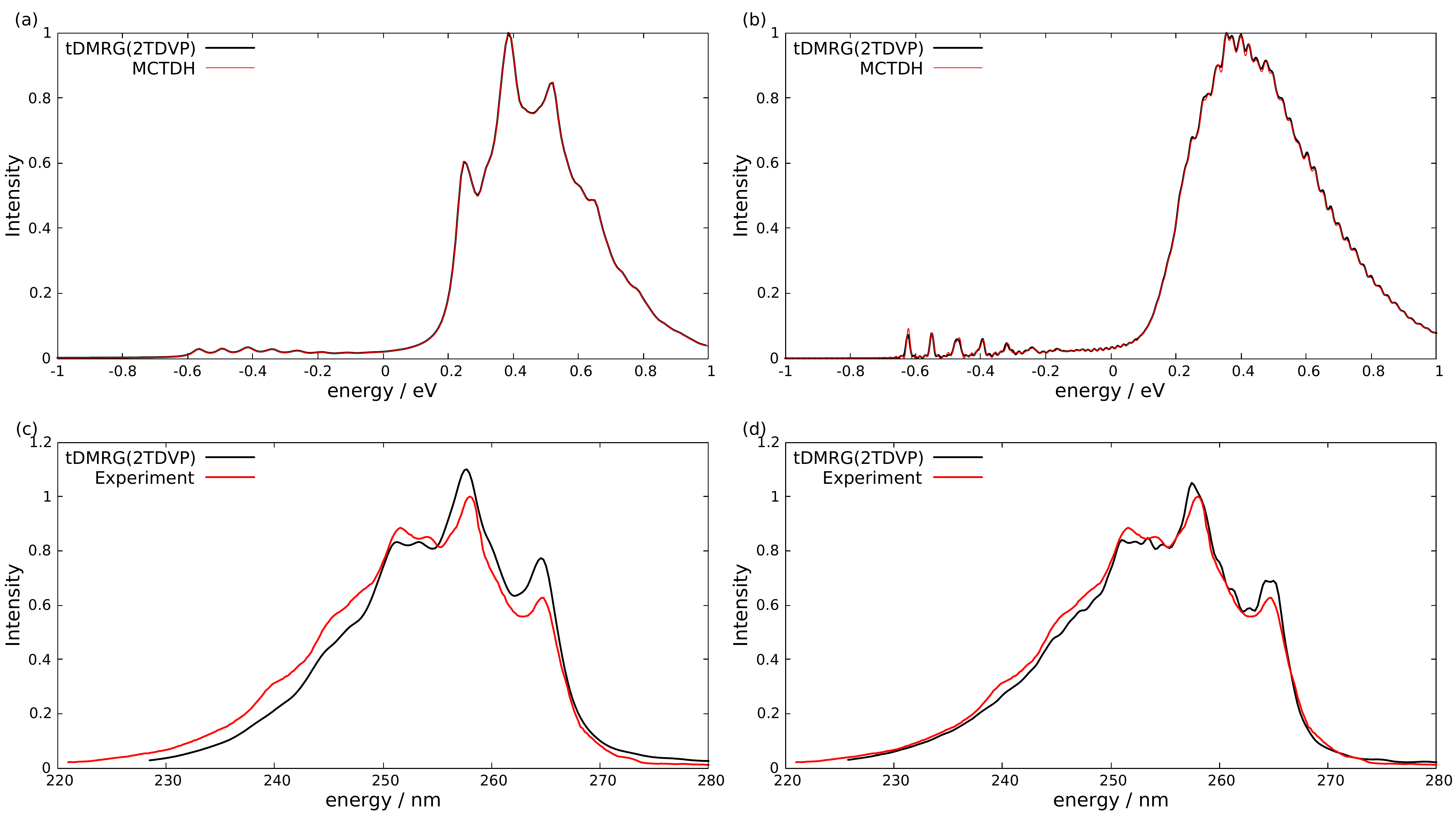}
\caption{Results for the spectra of the molecule pyrazine. (a) Spectrum (black line) of the 4-mode model with parameters in reference \cite{worth1996effect} comparing to MCTDH simulation (red line) result ($\tau=30$ fs). (b) Spectrum (black line) of the 24-mode model with parameters as given in reference \cite{worth1996effect}, compared to a MCTDH simulation (red line) result ($\tau=30$ fs). (c) Spectrum (black line) of the 4-modes model with parameters as in reference \cite{raab1999molecular} compared to experimental results~\cite{yamazaki1983intramolecular} (red line) result ($\tau=30$ fs). (d) Spectrum (black line) of the 24-mode model with parameters as in reference \cite{raab1999molecular} compared to experimental results~\cite{yamazaki1983intramolecular} (red line) result ($\tau=50$ fs).} \label{fig: pyrazine3}
\end{figure}

Based on the conclusions of the above tests, the parameters (timestep $\delta t = 20.0$ a.u., truncation threshold  $\varepsilon=1 \times 10^{-8}$) of the 2TDVP method were optimized to speed up the performance and the 2TDVP method was then applied to two 4/24-mode model systems \cite{worth1996effect, raab1999molecular} of the pyrazine molecule. Among the 24 vibrational modes in the pyrazine el-vib interaction model, only 4 modes are strongly correlated with electronic states and the other 20 ones are bath-like (2 intermediate and other 18 weak ones).\cite{worth1996effect,raab1999molecular} Also considering the too huge number of possible ordering combination for this 24-mode system, we try to find its guide from a more economic test of the ordering issue of only 6-mode system (4 strong-coupling modes and 2 intermediate-coupling bath-like modes). The results showed that the two bath-like modes are preferable to locate at one end of the lattice with the optimized ordering of the 4 strong modes ($\{v9a, v6a, \vert\mathrm{S}_1\rangle, v10a, \vert\mathrm{S}_2\rangle, v1\}$) as shown in the above paragraph. This can speed up by around 10 times compared to a default ordering (S$_1$, S$_2$, followed by Bosonic sites with the ascending order of the magnitude of frequency value). Considering the similar bath-like behavior of all 20 modes, we further expanded the optimized ordering from $4+2$ to $4+20$, i.e. we put other 20 modes at one end of the lattice with the appropriate ordering of the core of 4 modes and 2 electronic states and the ordering is $\{v9a, v6a, \vert\mathrm{S}_1\rangle, v10a, \vert\mathrm{S}_2\rangle, v1\}$ followed by other 20 modes in the order of frequency value. By calculating the Fourier transform of the product of the autocorrelation function $C(t)$ and a damping function $f(t)=e^{-\vert t\vert/\tau}$, the spectrum of the pyrazine system was obtained. The results for the spectrum are shown in Fig.~\ref{fig: pyrazine3}. The results of the linearly approximated models are in good agreement with the results of the MCTDH method \cite{worth1996effect} in Fig~\ref{fig: pyrazine3} (a, b) and the results for second-order models \cite{raab1999molecular} are comparable to experimental results \cite{yamazaki1983intramolecular}. This nice agreement indicates that tDMRG is a feasible and accurate method for describing the quantum dynamics of realistic chemical systems.

\begin{figure}
\includegraphics[scale=0.28]{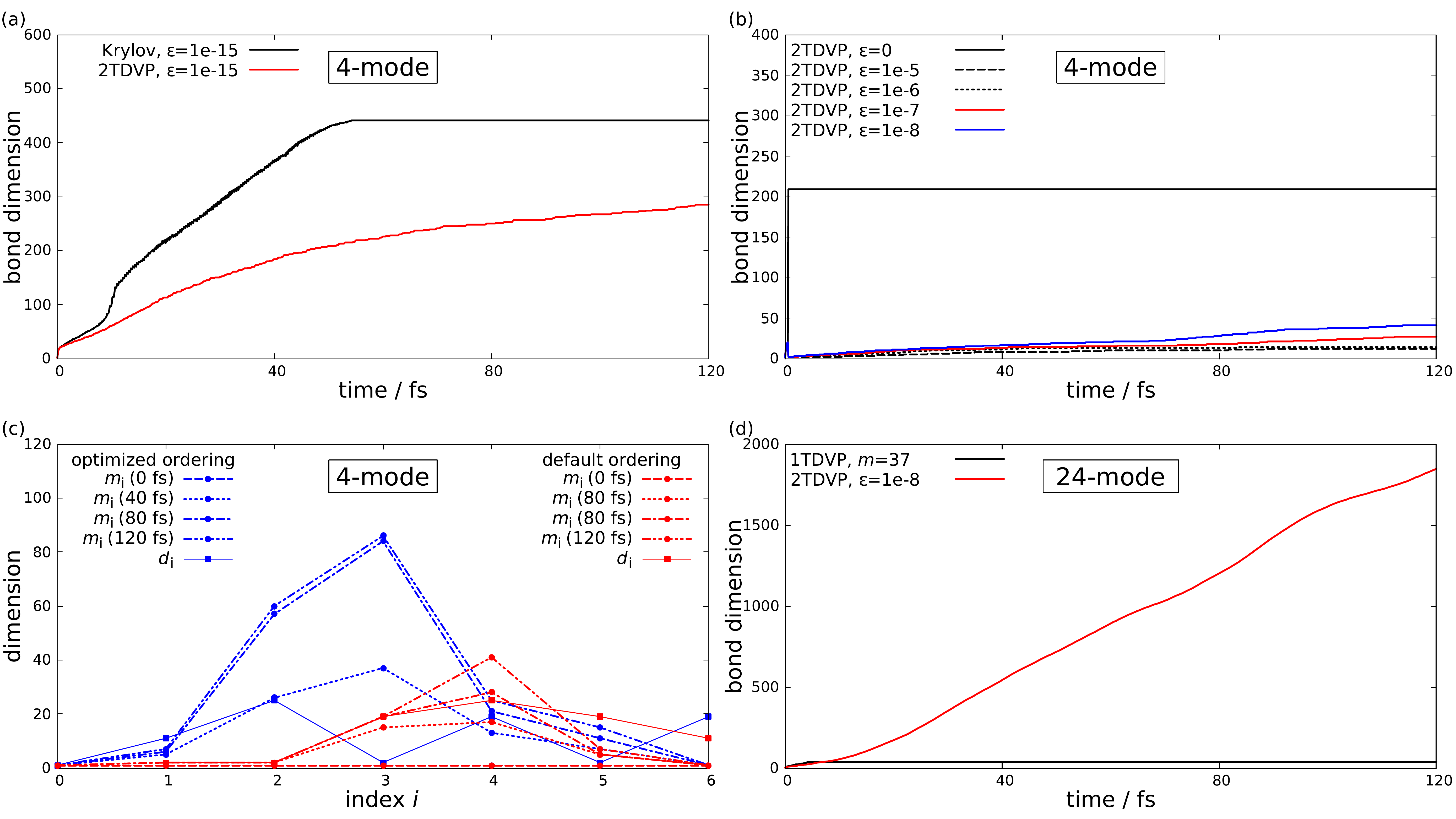}
\caption{Dynamics behaviour of bond dimension for the systems in Fig.~\ref{fig: pyrazine1}-\ref{fig: pyrazine3}. (a). Different methods testing corresponding with Fig. \ref{fig: pyrazine1} ($n_\mathrm{max}=20$), (b). Different truncations testing related to Fig.~\ref{fig: pyrazine2} (c, d), (c). Dimension of each bond legs (dash lines) and physical legs (solid lines) for the two ordering (red lines for default ordering and blue lines for optimized one) ordering tests and (d). the bond dimension of 24-mode system in Fig.~\ref{fig: pyrazine4} using 1TDVP and 2TDVP methods.} \label{fig: pyrazine4}
\end{figure}

As the bond dimension is directly related to the accuracy and efficiency of (t)DMRG, we further analyzed the time-evolution of MPS bond dimension. In Fig.~\ref{fig: pyrazine4}(a), from the results of the 4-mode system with $n_\mathrm{max}=20$ and $m_\mathrm{MPO}=4$ (corresponding to $m_\mathrm{limit}=410$), we could notice that both global Krylov and 2TDVP methods could increase the bond dimension adaptively. Besides, it is evident to find that 2TDVP method is more efficient than global Krylov method because much smaller bond dimensions are required for 2TDVP during the time evolution. For the performance of DBSS with different truncation thresholds, we also calculated the bond dimension dynamics in Fig.~\ref{fig: pyrazine4}(b) via 2TDVP methods corresponding to the system ($m_\mathrm{limit}=209$ and $m_\mathrm{MPO}=4$) in Fig.~\ref{fig: pyrazine2} (c, d). The blue line for converged simulation ($\varepsilon=1 \times 10^{-8}$) implies that efficient compression ($m=45$ at 120 fs compared with $m_\mathrm{limit}=209$) can be done by 2TDVP method without losing accuracy. This is also consistent with the converged result of 1TDVP method with $m=49$ in Fig.~\ref{fig: pyrazine2} (e, f). Unlike the small MPS bond dimension required for time-evolution of 4-mode system, the bond dimension of 24-mode system with $m_\mathrm{MPO}=14$ (shown in Fig.~\ref{fig: pyrazine3} (d)) grows very fast from 1 to around 1800 at 120 fs (see Fig.~\ref{fig: pyrazine4} (d)) using 2TDVP method. At the same time, 1TDVP fails to correctly describe the dynamics behavior of 24-mode system because its MPS bond dimension cannot increase to a sufficiently large number to approximate the evolved MPS. In addition, the detailed information of dimension of each bond leg allows us to estimate the wavefunction compression efficiency of MPS framework comparing to the MCTDH approach. Like the shape of the bond legs shown in Fig.\ref{fig: pyrazine4}(c), the dimension of the bond legs of MPS in the 24-mode system ($\varepsilon=1 \times 10^{-8}$) is hill-like in our 2TDVP calculation which means size of the wavefunction will be much smaller than that using the fixed bond dimension approach. Size of the MPS wavefunction in Fig.~\ref{fig: pyrazine4}(d) is time-dependent, being $212$, $1.3\times 10^7$, and $7.3\times 10^7$ at $t=0,\,60\,\mathrm{and}\,120$ fs respectively, and the time-independent values are $4.6\times 10^7$ and $4.5\times 10^5$ for MCTDH and ML-MCTDH individually.\cite{vendrell2011multilayer} Thus, data compression efficiency of MPS is similar with MCTDH while ML-MCTDH is more compressed considering the
complexity of the ML-MCTDH structure. Further testing of time consumption of 24-mode pyrazine system (120 fs) shows that our 2TDVP calculation ($\varepsilon=1 \times 10^{-8}$, CPU time: 10.3 h) is slightly faster than MCTDH (13.7 h) using single CPU core of Xeon 8163. Considering the only polynomial scaling of tDMRG's computational costs with respect to the system size, rather than the exponential scaling of MCTDH, tDMRG can be expected to deal properly with larger systems, e.g. the singlet fission case with 183 vibrational modes in the next subsection.

\subsection{Singlet fission in a molecular dimer} \label{sec: 3.2}
The singlet fission (SF) is a spin-allowed photophysical process that splits one singlet excitation state into two triplet excitons in various organic materials.\cite{smith2010singlet, smith2013recent, casanova2018theoretical} It is generally assumed that there are three crucial groups of electronic states for singlet fission, namely local excitation (LE) states, charge transfer (CT) states and triplet pair (TT) states. Many experimental and theoretical studies of SF have indicated that vibration modes play very important roles for SF \cite{teichen2012a, berkelbach2013microscopicI, berkelbach2013microscopicII, berkelbach2014microscopicIII, bakulin2016real, tamura2015first, yao2016coherent, fujihashi2017effect, morrison2017evidence, miyata2017coherent, zang2017quantum, tempelaar2017vibronicI, tempelaar2017vibronicIi, tempelaar2018vibronicIII, xie2019exciton}, but the full quantum treatment of both the electronic and vibrational part is still challenging because of the large configuration space of many electronic states and many vibration or phonon modes.

Here we test the tDMRG dynamics of the SF, comparing our results to a recent ML-MCTDH simulation by Lan et al. \cite{zheng2016ultrafast}. For this, we adopt the model used in ref. \cite{zheng2016ultrafast} which contains three electronic states (one LE state, one CT state and one TT state) and a linear local el-vib coupling approximation (i.e. $g^{IJ}_{ij}=0$ and $g^{I}_{ij}=0 (i\neq j)$). The parameters for the electronic part are listed in Tab~\ref{tab: el_SF}.

\begin{table}
\caption{The parameters of the SF electronic Hamiltonian matrix as in ref.~\cite{zheng2016ultrafast}. (unit: eV) }\label{tab: el_SF}
\begin{ruledtabular}
\begin{tabular}{cccc}
   &    LE &    TT &    CT \\
\hline
LE &  0.10 &  0.00 & -0.05 \\
TT &  0.00 &  0.00 & -0.05 \\
CT & -0.05 & -0.05 &  0.30 \\
\end{tabular}
\end{ruledtabular}
\end{table}

The vibration and el-vib interaction terms are characterized by the continuous Debye-type spectral density
\begin{equation}
J_i(\omega)=\frac{2\lambda\omega\omega_0}{\omega^2+\omega_0^2},
\end{equation}
where $\lambda=0.1$ eV represents the strength of the el-vib coupling and $\omega_0=0.18$ eV is the characteristic frequency of the bath. $i$ is the index for three electronic states. Consequently, the local el-vib couplings can be computed by discretizing the spectral density
\begin{equation}
g_{i}^{I}=\sqrt{\frac{2}{\pi}J_i(\omega_I)\Delta\omega_I}
\end{equation}
with a discrete series of bath frequencies $\{\omega_I\}$ in a given region and $\Delta\omega_I=\omega_I-\omega_{I-1}$. In the calculations here, we use 10, 8 and 6 as the the maximal occupation numbers for the vibrational modes coupled to LE, CT and TT states respectively after the convergence tests.

\begin{figure}
\includegraphics[scale=0.24]{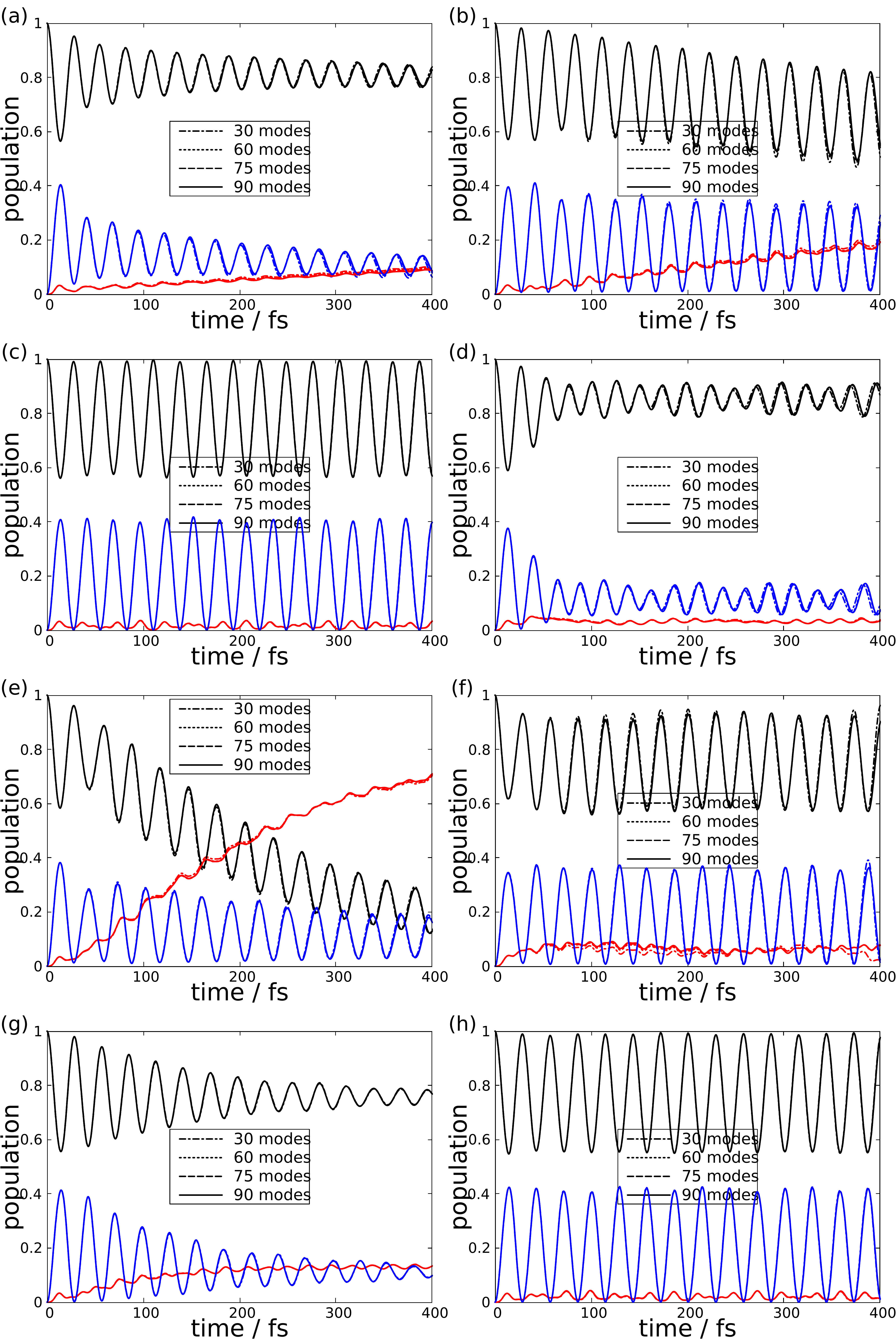}
\caption{The time-evolution for the populations of three electronic states (LE state with black lines, CT state with blue lines and TT state with red lines) for the SF dynamics with the bath in specific energy regions of (a) 0-0.075 eV (R1), (b) 0.075-0.108 eV (R2), (c) 0.108-0.125 eV (R3), (d) 0.125-0.165 eV (R4), (e) 0.165-0.2 eV (R5), (f) 0.2-0.3 eV (R6), (g) 0.3-0.35 eV (R7) and (f) 0.35-0.4 eV (R8). Different numbers of modes are used to test the convergence of the representation of the continuous vibration bath.} \label{fig: SF1}
\end{figure}

First, the relevant region of the spectral density (0-0.4 eV) is separated into eight regions following reference \cite{zheng2016ultrafast}, namely 0-0.075 eV (R1), 0.075-0.108 eV (R2), 0.108-0.125 eV (R3), 0.125-0.165 eV (R4), 0.165-0.2 eV (R5), 0.2-0.3 eV (R6), 0.3-0.35 eV (R7) and 0.35-0.4 eV (R8), to investigate the effects of bath modes in different energy ranges. The issue of bath representation convergence is tested by increasing the number of discrete vibration modes (10, 20, 25 and 30 modes for each electronic states in all eight regions). The results are illustrated in Fig.~\ref{fig: SF1}. The converged results of the electronic population evolution show good agreement with Lan et al.'s ML-MCTDH simulation. Our results with different numbers of discrete modes in regions R3, R4, R7 and R8 show very quick convergence while R1, R2, R5 and R6 require much larger number of vibration modes to converge the dynamics.

\begin{figure}
\includegraphics[scale=0.35]{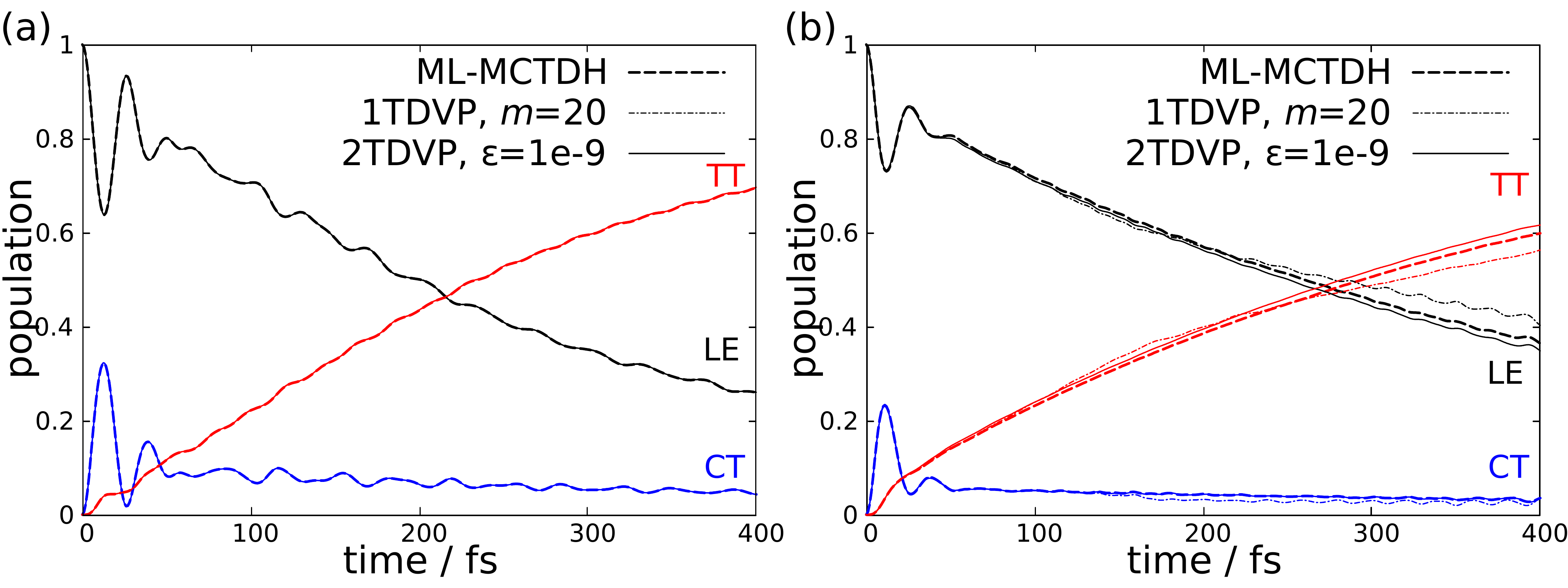}
\caption{The time-evolution for the populations of three electronic states of the SF dynamics coupled to the bath in the energy regions of (a) R4 + R5 + R7 (90 modes) and (b) 0-0.4 eV (183 modes) by tDMRG and ML-MCTDH \cite{zheng2016ultrafast}.} \label{fig: SF2}
\end{figure}

The population results in Fig~\ref{fig: SF1} clearly indicate that the phonon modes in energetic ranges R4, R5 and R7 play the most important roles influencing the SF dynamics. The coupling to the bath modes in regions R4 and R7 decreases the oscillation amplitudes of the LE and CT populations significantly (Fig~\ref{fig: SF1}(d, g)) while the TT state of SF is ultrafast formed mainly due to the effect of the coupling to the bath in range R5 (Fig~\ref{fig: SF1}(e)). Therefore, the dynamics of SF accounting for the effects of the bath may be simulated approximately by including the three vibration regions R4, R5 and R7. We also calculated the dynamics of the SF system in the case where all eight domains of the bath are included, compared to the reduced bath (R4, R5, R7).\cite{zheng2016ultrafast} Our population dynamics results by 1TDVP and 2TDVP methods as displayed in Fig.~\ref{fig: SF2}(a) are in very good agreement with the ML-MCTDH simulation in ref. \cite{zheng2016ultrafast} for the reduced bath coupled system (90 modes, $m=30$ for 2TDVP at $t=400$ fs), showing again that tDMRG can be used for the accurate and efficient simulation of quantum dynamics in large chemical systems. But it is also worth to mention that, according to Fig.~\ref{fig: SF2}(b), 1TDVP has larger quantitative deviations from ML-MCTDH than 2TDVP result for the full window (183 modes, $m=450$ for 2TDVP at $t=400$ fs).

\begin{figure}
\includegraphics[scale=0.24]{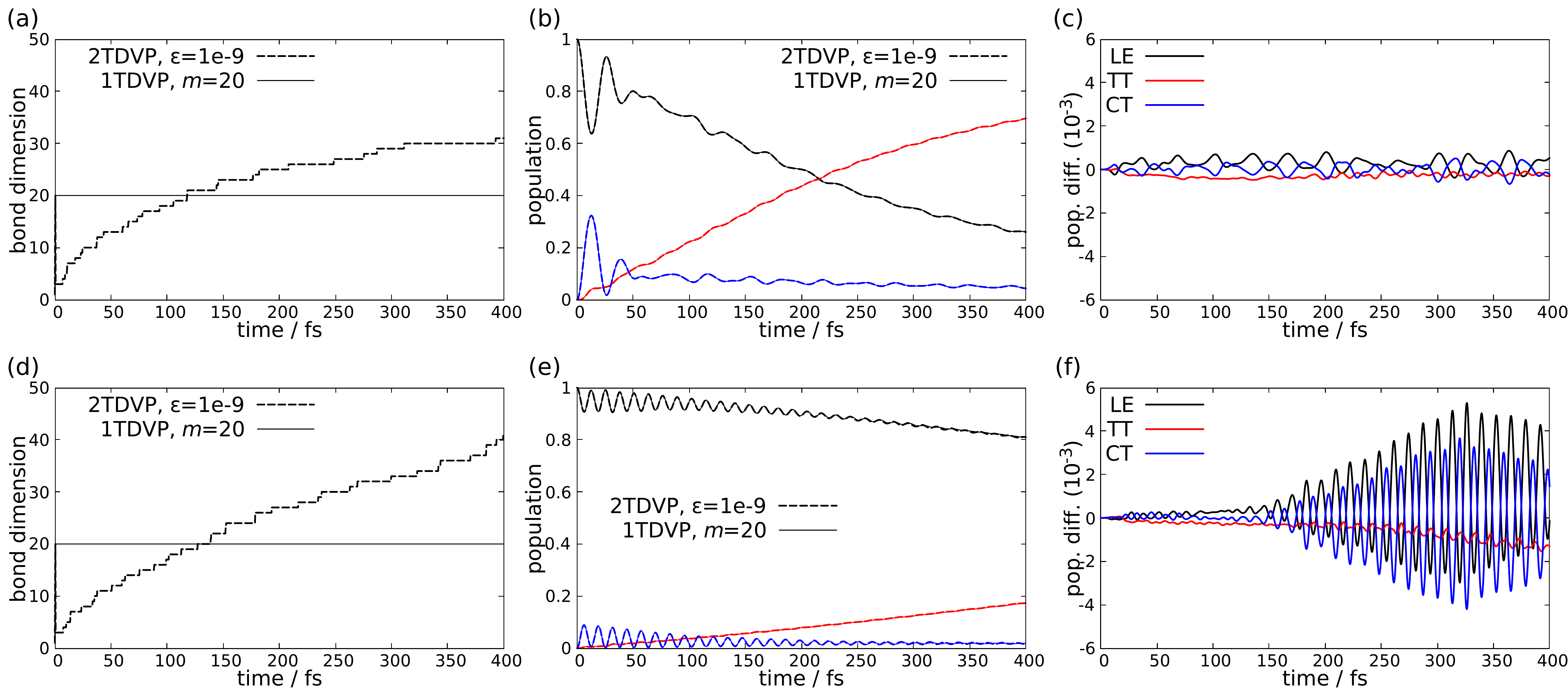}
\caption{Dynamics results of bond dimension and population of electronic states for SF systems with different CT energies via 1TDVP/2TDVP methods. (a, d) the bond dimension for models with $E_{\mathrm{CT}}=$0.3 and 0.5 eV respectively, (b, e) populations of electronic states for the two model systems and (c, f) the difference of populations between 2TDVP and 1TDVP for the two model systems.} \label{fig: SF3}
\end{figure}

\begin{figure}
\includegraphics[scale=0.24]{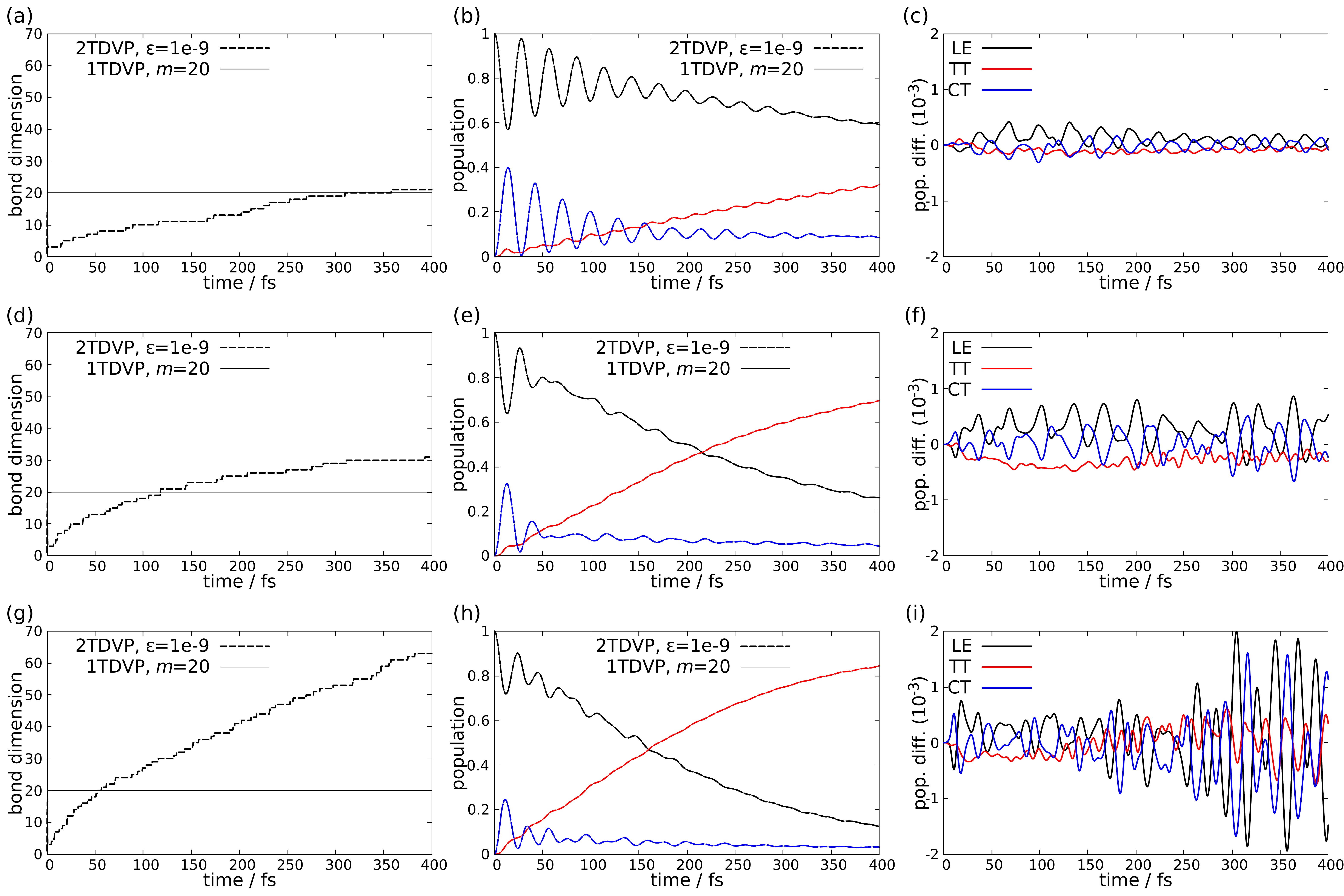}
\caption{Dynamics results of bond dimension and population of electronic states for SF systems with different strength of el-vib coupling via 1TDVP/2TDVP methods. (a, d, g) the bond dimension for models with $\lambda=$0.05, 0.1 and 0.15 eV respectively, (b, e, h) populations of electronic states for the three model systems and (c, f, i) the difference of populations between 2TDVP and 1TDVP for the three model systems.} \label{fig: SF4}
\end{figure}

Furthermore, we tested the tDMRG performance with respect to parameter values of model in the Hamiltonian by comparing 1TDVP/2TDVP calculation of the model systems with different parameters (energy of CT state, strength of el-ph coupling). The results of the population difference of 1TDVP/2TDVP calculation as well as the dynamics of bond dimension were shown in Fig.\ref{fig: SF3} and \ref{fig: SF4}. In general, for our tests of 90-mode systems, 1TDVP can give reasonable simulation results, which are comparable to 2TDVP ones. By enlarging the energy gap between CT state and other stats, singlet fission rate decreases and bond dimension of MPS in 2TDVP (Fig.~\ref{fig: SF3} (d)) becomes larger than original model in Fig.~\ref{fig: SF3} (a), leading to the increasing difference between 1TDVP and 2TDVP population results from 1$\times 10^{-3}$ to 6$\times 10^{-3}$. In Fig.~\ref{fig: SF3} (e), one can also notice that 1TDVP calculation underestimate the CT/LE population oscillation caused by the electronic coherence between LE and CT states in the long time limit after 200 fs. In Fig.~\ref{fig: SF4}, we also consider a weaker and a stronger el-ph coupling regimes by tuning the $\lambda$ value ($\lambda=0.05$, $0.1$ and $0.15$ eV respectively). It is found that, upon increasing the el-ph coupling strength, singlet fission rate increases and the MPS bond dimension for 2TDVP increases more rapidly. At the same time, the population difference between 1TDVP and 2TDVP results increases from 5$\times 10^{-4}$ to 2$\times 10^{-3}$.

\section{Summary} \label{sec: 4}
In order to validate the accuracy and efficiency of various tDMRG methods for realistic electron-vibration/phonon systems, we benchmarked the tDMRG calculations with different algorithms (global Taylor, global Krylov, local 1TDVP and 2TDVP) for the S$_1$/S$_2$ internal conversion in the pyrazine molecule and applied a TDVP calculation to the singlet fission in a molecular dimer. Our tDMRG results were compared with (ML-)MCTDH and experimental results.

We find that time-evolution of large systems via 2nd-order Taylor expansion may crash due to its large truncation error of time evolution propagator $\hat{U}$ with respect to the system size while local methods (e.g. TDVP methods) give reasonable results. Among the two TDVP variants (one- and two-site TDVP), 2TDVP is found to be able to give reasonable results for all our tested systems. On the other hand, 1TDVP has a significant advantage of computational efficiency and also works well for simple systems (4-mode pyrazine system and 90-mode SF system), which usually only requires a small MPS bond dimension around a few tens. But 1TDVP can hardly achieve quantitative or even qualitative correct results for complicated large systems (24-mode pyrazine system and 183-mode SF system) unless an MPS with sufficiently large bond dimension (around a few hundreds or more) is preliminarily prepared. Another very important advantage of 2TDVP over 1TDVP will appear in conjunction with the use of good quantum numbers, as it allows for the resizing of the size of blocks of states with the same quantum numbers, adapting to the changing system, while 1TDVP does not. But this advantage will only come to play in scenarios with more electronic states, not the studies in this work with only two or three electronic states.

Several key parameters in the tDMRG calculation including the truncation error threshold, time interval and ordering of local sites were investigated to strike the balance between efficiency and accuracy of results. It is worthwhile to emphasize that the ordering of sites will be a very important factor influencing the efficiency of TDVP methods. We suggest to arrange strongly correlated sites close to each other and locate the entangled sites with small basis at the center of the chain.

The comparison of dynamics results of optimized tDMRG simulations to the benchmarks of (ML-)MCTDH and experimental results for both systems confirms that the tDMRG methods (particularly, the 2TDVP method) are good candidates for quantum dynamics calculations for chemical systems. Our tests provide guidelines for future applications of tDMRG methods to simulate quantum dynamics of realistic chemical systems.

\section{Acknowledgments}
The work was supported by the National Natural Science
Foundation of China (Grant Nos. 21722302 and 21673109) and by the Deutsche Forschungsgemeinschaft (DFG, German Research Foundation) under Germanys Excellence Strategy--EXC-2111—No. 390814868. We thank Victor S. Batista, Claudius Hubig, Zhenggang Lan, Sam Mardazad, and Yu Xie for helpful discussions.

\bibliography{ref}

\end{document}